\documentclass{elsart}

\usepackage{natbib}

\usepackage{graphicx}

\usepackage{amssymb}

\begin{document}

\begin{frontmatter}



\title{Localization of  Multi-State Quantum Walk\\ in One Dimension}

\author{Norio Inui}
\address{Graduate School of Engineering, 
University of Hyogo, \\
2167, Shosha, Himeji, Hyogo, 671-2201, Japan}
{\small E-mail: inui@eng.u-hyogo.ac.jp}

\author{Norio Konno}
\address{Department of Applied Mathematics, 
Yokohama National University,  \\
79-5 Tokiwadai, Yokohama, 240-8501, Japan}
{\small E-mail: norio@mathlab.sci.ynu.ac.jp}

\begin{abstract}
Particle trapping in  multi-state quantum walk on a circle is studied. 
The time-averaged probability distribution of a particle which moves four different lattice sites according to four internal states is
calculated exactly.   In contrast with  ``Hadamard walk" with only two internal states, the particle remains at the initial position with high probability.
The time-averaged probability of finding the particle decreases exponentially
as distance from a center of a spike. This implies that the  particle is trapped in a narrow region.
This striking difference is minutely explained from difference between  degeneracy of eigenvalues of the time-evolution matrices.
The dependence of the particle distribution on  initial conditions is also considered.\end{abstract}


\end{frontmatter}

\section{Introduction}

The quantum walks attract our attention not only in 
a fundamental field of quantum mechanics 
[1] but also in a filed of quantum information [2]
(see also recent review with many references [3]).
This is because the quantum walks will be achieved on quantum computers [4], and it 
may be used as a new quantum algorithm.

A classical random walker on a line moves to a nearest left site  {\it or} a nearest right site randomly at every time step. 
On the other hand, a quantum walker moves, in superposition,  both left {\it and} right according to additional degree of freedom called
``chirality". First of all, we describe the Hardamrd walk, which is the most well-known quantum walk.
The Hadamard walk is characterized by two parameters. The first is the spatial position and  the second is the chirality, which takes
values ``$+1$" and  ``$-1$". The corresponding Hilbert spaces are $H=\bf C \rm ^{\infty}$  and $H_{I}=\bf C \rm^{2}$,
where $\bf C \rm$ is the set of complex numbers. 
The position space consists of basis states $|x \rangle \in H (-\infty < x < \infty)$ corresponding to the
walker located at $n$-th lattice site.
Possible experimental implementations of the Hadamard walk have been proposed by a number of authors
[4-6].
For example, a quantum walker
is an atom and  internal states of the atom such as hyperfine structure are used to distinguish the chirality.

Let  $\psi(s,x,t)$  be  a value of
wave function at  position $x$ and time $t$ with the chirality $s \in \{ -1,1 \}$. The wave function is  changed
by superimposing $\psi(-s,x-s,t-1)$ on $\psi(s,x-s,t-1)$ at ever time step (details are defined in the next section). 
Assume that a walker exists  at the origin in spatial-temporal plane, then we find the walker in the symmetrical Hadamard  walk with
the same probability at the position 1 or $-1$  after a single transformation. This behavior is the same as symmetric classical random walk.
The primary question in the quantum walk is to determine the probability distribution of the walker after repeating the transformation of wave function.
The recent extensive studies  made clear mathematical properties of  the Hadamard walk. We know that the probability distribution
is quite different from that of classical random walk. When the initial wave function is given by
$\psi(-1,0,0)=1/2$ and $\psi(1,0,0)=i/2$, (symmetric case),  we see that double peaks spread in the mutually different direction.
Furthermore the standard deviation increases in proportion to time. This means that the quantum walker spreads more quickly 
than classical random walker, and  it is a desirable feature for seeking problems.

Recent studies  revealed that there are two distinguishing profiles in modified Hadamard walks on an infinite space.
When the walker exists initially at the origin, the one of shapes is concave in a broad way and the other shape is a spike.
The spike at the initial time in the symmetrical Hadamard walk sprits two peaks and they move away from the origin as time increases.
The probability of finding the walker near the origin decreases and approaches to zero.
Thus the probability distribution of the original Hadamard walk belongs to the former.
On the other hand, we can always find a walker with high probability near the origin in the later case.
We know little about the spiky probability distribution in the later case.  
The existence of the spikes are already reported in the two-dimensional Grover walk
[7,8], one-dimensional quantum walk as
many coins [9], but, no probability distributions are calculated explicitly.  In this paper we  consider quantum walk 
with four internal states, 
which is the same  as the quantum walk with many coins essentially, and compute analytically the probability distribution. 
The Hadamard walker described above has two internal quantum states
and moves to only nearest neighbor sites, while a walker in  ``four-state quantum walk"  can move beyond nearest neighbor sites  by a single
transformation.  

The rest of this paper is organized as follows. We define multi-state quantum walk generally and  show a result obtained by simulation in the next section. 
In section 3, eigenvalues of the time evolution matrix are exactly calculated and  the wavefunction is expressed using those eigenvalues.
In section 4, we introduce  a time-averaged probability and show that the quantum walker is found with high probability at the initial position.
In section 5, a quantum walker with three internal states, in which a walker jumps asymmetrically, is considered and it is shown that
localization is not observed.   


\section{Multi-state quantum walk}
 We consider  a quantum walker with $n$ different internal states and moves to $n$ different lattice sites according to
the internal states. Assume that $n$ is an even integer larger than 2. Let
 $s \in \Gamma \equiv \{\pm1,\pm2,\cdots, \pm n/2 \}$. We then define the transformation of
$n$-state quantum walk by
\begin{eqnarray}
\psi(s,x,t+1) = \sum_{s' \in \Gamma} \delta_{s,s'} \psi(s',x-s,t),
\label{eqn:geT}
\end{eqnarray}
where $\delta_{s,s' \neq s}=2/n$ and $\delta_{s,s}=2/n-1$. 
This transformation is quite simple, but, it produces rich quantum effects.
If the number of internal state is four, we regard the above transformation as the quantum version of the 
classical random walk in which a walker jumps to each nearest site and next nearest site with the same probability $1/4$.

Our main conclusion arguing that the particle in $n$-state quantum walk is trapped near the starting point is available
for any even $n$ except $n=2$, but  in this paper we focus on four-state quantum walk to avoid excessive generality.
The transformation with $n=2$ is diffrent from the Hadamard transformation, and the dynamics of $2$-state quantum walk
is trivial.

We begin to calculate the probability distribution of the quantum walks on a circle containing
$N$ sites [10]. We define the probability  of finding a walker at position $x$ and
time $t$  by
\begin{eqnarray}
P(x,t) \equiv \sum_{s \in \Gamma} \psi(s,x,t)(\psi(s,x,t))^{\ast}, 
\label{eqn:Prbex}
\end{eqnarray}
where ``$\ast$" denotes complex conjugate. To compare the probability distribution between the Hadamard walk and four-state quantum walk, 
Fig.1 shows the snapshot of  probability distribution at $t=50$  starting from an initial wave function:
\begin{eqnarray}
\psi(s,x,0) &=&
\left \{
\begin{array}{ll}
\hspace{5mm}\frac{1}{\sqrt{n}} \hspace{10mm} & x=0 \,\, \mbox{and} \,\, s<0, \\
\hspace{5mm}\frac{i}{\sqrt{n}} & x=0 \,\, \mbox{and} \,\, s>0, \\
\hspace{5mm}0  & x \neq 0.
\end{array}
\right. 
\label{eqn:iniwf}
\end{eqnarray}

\begin{figure}[h]
\centering
\includegraphics{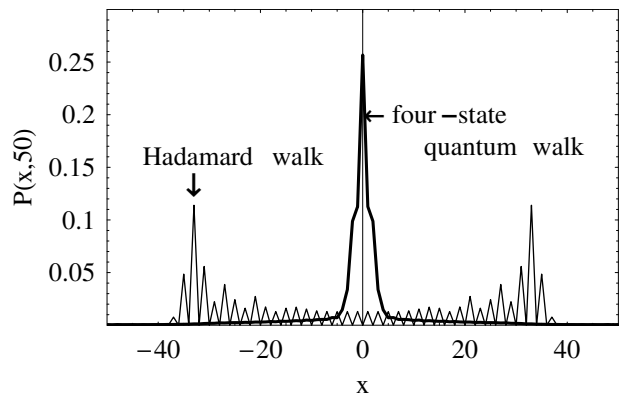}
\caption{\label{fig:Figure2} 
Comparison of probability distributions between the Hadamard walk and four-state quantum walk at $t=50$.
A walker initially exists  at the origin in both cases.  The walker in the Hadamard walk moves away form the origin.
The walker in four-state quantum walk, by contrast, stays at the origin with high probability.
}
\end{figure}
 In contrast with the Hadamard walk we clearly find a single spike
at the origin in  four-state quantum walk. A similar spike has  also been  observed in two-dimensional quantum walk by
simulation [7] and the height of time-averaged probability at the origin is exactly calculated [8]. However, 
the probability distribution  has not been obtained explicitly as a function of the location.
For these reason, the main purpose of this paper is
to calculate the time-averaged probability distribution of four-state quantum walk and show that the spike in Fig.1 does not
disappear even in the thermodynamic limit.

\section{Transfer matrix and eigenvalues}

The time evolution of the wavefunction of the multi-state quantum walk $\Psi(t)$ is determined by transformation Eq. (\ref{eqn:geT}) and
it is expressed briefly by introducing a time evolution operator $M$ satisfying $\Psi(t+1)=M \Psi(t)$. Since the matrix $M$ is unitary, 
$\Psi(t)$ is explicitly given by transforming $M$ to a diagonal matrix as a function of the time. 
Using Fourier transformation  it is found that 
 eigenvalues of $M$ are calculated  from the eigenvalues of the following matrix:
\begin{eqnarray}
H_{k}=
\frac{1}{2}
\left[
\begin{array}{cccc}
\omega^{-2k} &     0        &     0         & 0 \\
0           &  \omega^{-k}  &     0         & 0 \\
0           &     0        &  \omega^{k}  & 0 \\
0           &     0        &     0         &  \omega^{2k} 
\end{array}
\right]
\left[
\begin{array}{cccc}
-1 &  1   &   1  &  1 \\
1  & -1   &   1  &  1 \\
1  &  1   &  -1  &  1 \\
1  &  1   &   1  & -1 
\end{array}
\right],
\label{H}
\end{eqnarray}
where $\omega=e^{2 \pi i/N}$.
Changing the value of $k$ from 0 to $N-1$, we get all eigenvalues of $M$ (see details in Ref. [11]).
The eigenvalues of $H_{k}$ are given by
\begin{eqnarray}
\lambda_{k,1} &=& -1, \hspace{28mm} \lambda_{k,2} =  1, \\
\lambda_{k,3} &=&  \frac{-g_{k} - i \sqrt{4 - g_{k}^2}}{2}, \hspace{5mm}
\lambda_{k,4} =  \frac{-g_{k} + i \sqrt{4 - g_{k}^2}}{2},  
\label{eqn:lambda}
\end{eqnarray}
where $g_{k} = \cos \theta_{2k} + \cos \theta_{4k}$ and $\theta_{k}=k \pi/N$ for $k=0,1,\cdots, N-1$.
We stress here that the eigenvalues $-1$ and $1$ are independent of value $k$, put simply, they are strongly degenerative.
In fact, the degree of degeneracy for $-1$ and $1$  are $N+2$ and $N$, respectively.  Since 
$g_{k}=g_{N-k}$ for $k>0$, the  degree of degeneracy of other eigenvalues is two unless $N$ is a multiple of 5.
In case of $N=even$, the classification of eigenvalues is more complicated. Thus we suppose that $N$ is odd 
and $N$ is not a multiple of 5 in this paper.

All eigenvalues are now obtained and then the wave function is formally expressed  by
\begin{eqnarray}
\psi_{N}(s,x,t) &=&c_{s,N}^{(1)}(x)(-1)^{t}+c_{s,N}^{(2)}(x) \nonumber \\
&&\hspace{5mm}+\sum_{k=1}^{(N-1)/2} c_{s,k,3,N}(x)e^{i \omega_{k,3} t}+ \sum_{k=1}^{(N-1)/2} c_{s,k,4,N}(x)e^{i \omega_{k,4} t},
\label{eqn:wfformal}
\end{eqnarray}
where $\omega_{k,n}$ denotes the argument of $\lambda_{k,n}$, 
and the first and second terms express the contribution of eigenvalue 
$-1$ and 1 to the wave function. 
The coefficients in the above formula are calculated from the eigenvectors of $M$.
Let $\phi_{k,n}$ be the eigenvectors corresponding to the eigenvalue $\lambda_{k,n}$ of $H_{k}$  
and $\phi_{k,n,j}$  be the $j$-th element for $i=0,1,2,3$. We assume additionally $\{ \phi_{k,1},\cdots,\phi_{k,4} \}$ is an orthonormal basis. 
Then the $i$-th element of eigenvectors of the matrix $M$ with $\lambda_{k,n}$, which becomes
an orthonormal basis is obtained by
\begin{eqnarray}
\Phi_{k,n,j} &=& \frac{1}{\sqrt{N}}\phi_{k,n,j \,\mbox{\tiny mod}\, 4} \omega^{k \lfloor j/4 \rfloor},
\label{eqn:othegienV}
\end{eqnarray}
where $\lfloor z \rfloor$ denotes the integer part of $z$. 
If the walker exists initially at the origin, 
the coefficient $c_{s,k,3,N}(x)$, for example, is expressed by
\begin{eqnarray}
\hspace{-5mm}c_{s,k,3,N}(x) =\frac{1}{N}\sum_{j=1}^{4}\{ \omega^{kx} \phi_{k,3,s} (\phi_{k,3,j})^{\ast} +\omega^{(N-k)x} \phi_{N-k,3,s} 
(\phi_{N-k,3,j})^{\ast}  \}\psi(s_{j},0,0), \nonumber\\
\label{eqn:csk3}
\end{eqnarray}
where $s_{1}=-2, s_{2}=-1, s_{3}=1$, and $s_{4}=2$.


\section{Time-averaged probability}
In general, the value of wave function at any fixed position $x$ does not converge to a constant value. Thus we here define the time-averaged probability of finding the particle at the position
$x$ and in the internal state $s$  in order to show the
localization in long time scale  as
\begin{eqnarray}
\bar{P}_{N}(s,x) &=& \lim_{T \rightarrow \infty} \frac{1}{T} \sum_{t=0}^{T-1} \psi_{N}(s,x,t)(\psi_{N}(s,x,t))^{\ast}.
\label{eqn:Pavs}
\end{eqnarray}
Noting that  $\lim_{T \rightarrow \infty} \sum_{t=0}^{T-1} e^{i( \omega-\omega')t}/T=0$ for $\omega \neq \omega'$,
we have the following equation:
\begin{eqnarray}
\hspace{-5mm}\bar{P}_{N}(s,x) &=&  \left|c_{s,N}^{(1)}(x) \right|^{2}+\left|c_{s,N}^{(2)}(x) \right|^{2}+\sum_{k=1}^{(N-1)/2} (|c_{s,k,3,N}(x)|^{2}+|c_{s,k,4,N}(x)|^{2}). \hspace{3mm}
\label{eqn:Pavs2}
\end{eqnarray}

Let us consider time-averaged probability in the limit $N \rightarrow \infty$ below. Combining Eqs. (\ref{eqn:othegienV}) and (\ref{eqn:csk3}), we 
find that $|c_{s,k,3,N}|$ and $|c_{s,k,4,N}|$ decrease in the form $1/N$ for large $N$. Therefore the third and fourth terms vanish in the limit of $N \rightarrow \infty$.
As a result, time-averaged probability on a circle containing infinite sites is simply given by
\begin{eqnarray}
\bar{P}_{\infty}(s,x)=\lim_{N \rightarrow \infty} \left (  \left|c_{s,N}^{(1)}(x)\right|^{2}+ \left|c_{s,N}^{(2)}(x)\right|^{2} \right).
\label{eqn:Pavinf}
\end{eqnarray}

Assume that the particle exists at the origin initially. Then  the values of $c_{s,\infty}^{(l)}(x)$ for $l=1,2$ 
in the right-hand side are determined from only four components of the initial state and they  are expressed in the following form:
\begin{eqnarray}
\left[
\begin{array}{c}
c_{-2,\infty}^{(l)}(x) \\
c_{-1,\infty}^{(l)}(x) \\
c_{1,\infty}^{(l)}(x) \\
c_{2,\infty}^{(l)}(x) \\
\end{array}
\right]
=
\left[
\begin{array}{cccc}
U_{-2,-2}^{(l)}(x)  & U_{-2,-1}^{(l)}(x) & U_{-2,1}^{(l)}(x) & U_{-2,2}^{(l)}(x) \\
U_{-1,-2}^{(l)}(x)  & U_{-1,-1}^{(l)}(x) & U_{-1,1}^{(l)}(x) & U_{-1,2}^{(l)}(x) \\
U_{1,-2}^{(l)}(x)  & U_{1,-1}^{(l)}(x) & U_{1,1}^{(l)}(x) & U_{1,2}^{(l)}(x) \\
U_{2,-2}^{(l)}(x)  & U_{2,-1}^{(l)}(x) & U_{2,1}^{(l)}(x) & U_{2,2}^{(l)}(x) \\
\end{array}
\right]
\left[
\begin{array}{c}
\psi(-2,0,0) \\
\psi(-1,0,0) \\
\psi(1,0,0) \\
\psi(2,0,0) \\
\end{array}
\right].
\nonumber
\\
\label{eqn:cmat}
\end{eqnarray}
In this formula, $U_{s,i}^{(l)}$ expresses a contribution from the initial state $\psi(s_{i},0,0)$ to $c_{s,\infty}^{(l)}(x)$.
If each component of the above matrix $U^{(l)}$ in Eq.(\ref{eqn:cmat})  is exactly obtained, then 
the time-averaged probability is immediately calculated from Eq.(\ref{eqn:Pavinf}) for any initial state.

Let us  now calculate $U_{s,i}^{(l)}(x)$. 
Using the set of eigenvectors of $H_{k}$,
$U_{s,i}^{(l)}(x)$ is given by 
\begin{eqnarray}
U_{s,i}^{(l)}(x)       & = &  \lim_{N \rightarrow \infty} \frac{1}{N}\sum_{k=0}^{N-1} \omega^{kx} \phi_{k,l,s} (\phi_{k,l,i})^{\ast}.
\label{eqn:Ud}
\end{eqnarray}
We use  the following orthonormal eigenvectors corresponding to $\lambda_{k,l}$  for $l=1,2$ and $k>0$:
\begin{eqnarray}
\phi_{k,l} =
\frac{1}{f_{l}(\theta_{k})}
\left [
\begin{array}{c}
\xi_{l,1}(\theta_{k}) e^{i 2 \theta_{k}} \\
\xi_{l,2}(\theta_{k}) e^{i 3\theta_{k}} \\
\xi_{l,3}(\theta_{k}) e^{i 5\theta_{k}} \\
\xi_{l,4}(\theta_{k}) e^{i 6\theta_{k}}
\end{array}
\right],  
\end{eqnarray}
where 
\begin{eqnarray}
f_{1}(\theta) &=& \sqrt{6+4\cos 2\theta},  \hspace{24.5mm} \nonumber \\
f_{2}(\theta) &=& \sqrt{2(1+\cos 2 \theta )(2+\cos 2 \theta +\cos 4 \theta )}, \nonumber \\
\xi_{1,1}(\theta_{k}) &=& -\xi_{1,4}(\theta_{k}) = 1, \nonumber  \\
\xi_{1,2}(\theta_{k}) &=& -\xi_{1,3}(\theta_{k})= 2\cos\theta_{k}, \nonumber  \\
\xi_{2,1}(\theta_{k}) &=& \xi_{2,4}(\theta_{k})  = 1+\cos(2\theta_{k}), \nonumber  \\
\xi_{2,2}(\theta_{k}) &=& \xi_{2,3}(\theta_{k})= \cos \theta_{k}+ \cos(3 \theta_{k}).  
\label{eqn:vn}
\end{eqnarray}

In case of $k=0$, since eigenvalues  are $\{-1,1,-1,-1\}$, and the value $-1$ degenerate,
we use the following  eigenvectors which are orthonormal basis
\begin{eqnarray}
\phi_{0,1} &=&  \left (-\frac{1}{\sqrt{2}},0,0,\frac{1}{\sqrt{2}} \right ), \nonumber \\
\phi_{0,2} &=&  \left (\frac{1}{2},\frac{1}{2}, \frac{1}{2},\frac{1}{2} \right ), \nonumber \\
\phi_{0,3} &=&  \left (-\frac{1}{\sqrt{6}},\frac{\sqrt{2}}{\sqrt{6}},0,-\frac{1}{\sqrt{6}} \right), \nonumber \\
\phi_{0,4} &=&  \left (-\frac{1}{2\sqrt{3}},-\frac{1}{2\sqrt{3}},-\frac{\sqrt{3}}{2},-\frac{1}{2\sqrt{3}} \right). 
\end{eqnarray}

Plugging above eigenvectors into Eq. (\ref{eqn:othegienV}) and carrying  out the summation
in Eq. (\ref{eqn:Ud}) we  have $U_{s,i}^{(l)}(x)$ exactly after a somewhat tedious calculation as follows:
\begin{eqnarray}
U^{(l)}_{11}(x) &=& g^{(l)}_{1}(|x|), \nonumber\\
U^{(l)}_{12}(x) &=&
\left \{
\begin{array}{ll}
g^{(l)}_{2}(x)   & \hspace{10mm} x \geq 1, \nonumber \\
g^{(l)}_{2}(1-x) & \hspace{10mm} x < 1,
\end{array}
\right. \\
U^{(l)}_{22}(x) &=&
\left \{
\begin{array}{ll}
g^{(l)}_{3}       & \hspace{10mm}x=0,  \\
-g^{(l)}_{1}(|x|) & \hspace{10mm} \mbox{otherwise}.
\end{array}
\right .
\label{eqn:U}
\end{eqnarray}
where
\begin{eqnarray}
&&g^{(1)}_{1}(x)= \frac{\alpha^{x}}{2 \sqrt{5}}, \hspace{5mm}
g^{(1)}_{2}(x) =-\frac{5+\sqrt{5}}{20} \alpha^{x},  \hspace{5mm} 
g^{(1)}_{3}   = \frac{5-\sqrt{5}}{10}, \nonumber\\ 
&&g^{(2)}_{1}(x)= \frac{\beta^{x}\cos (\gamma x+\delta_{1})}{2^{\frac{1}{4}}\,{\sqrt{7}}},   \hspace{5mm}
g^{(2)}_{2}(x)=-\frac{{\beta }^{x}\cos (\gamma x+\delta_{2})}{2^{\frac{1}{4}}\sqrt{7}\sqrt{-\beta}},  \nonumber\\
&&g^{(2)}_{3}   = \frac{1}{2} - \frac{{\sqrt{14 + 28\,{\sqrt{2}}}}}{28}.
\label{eqn:gG}
\end{eqnarray}
The value of $\alpha$, $\beta$, $\gamma$, $\delta_{1}$ and $\delta_{2}$ are given by

\begin{eqnarray}
\alpha &=&  \frac{-3 + {\sqrt{5}}}{2},   \hspace{5mm}
\beta = -\frac{1}{2}  - \frac{1}{{\sqrt{2}}} +\frac{{\sqrt{-1 + 2\,{\sqrt{2}}}}}{2}, \hspace{5mm}
\gamma = \tan^{-1} \left({\sqrt{11 + 8\,{\sqrt{2}}}} \right) \nonumber\\
\delta_{1} &=& \frac{\tan^{-1} {\sqrt{7}}}{2}, \hspace{5mm}
\delta_{2} = -\tan^{-1} \left ({\sqrt{3 + 2\,{\sqrt{2}} - 
       \frac{2\,{\sqrt{206 + 148\,{\sqrt{2}}}}}{7}}} \right).
\end{eqnarray}

When the number of size $N$ is large, the valuable $\theta_{k}=k \pi/N$ is almost continuous. Thus
the infinite summation is calculated by transforming the summation into the complex integration on the unit circle. The integration
is exactly calculated by residue theorem, and  
$\alpha$ and $\beta$ are poles  of the function of  $\omega^{kx} \phi_{k,l,s} (\phi_{k,l,i})^{\ast}$
in Eq. (\ref{eqn:Ud}), in which $\theta_{k}$ is replaced as continuous valuable. Other components are obtained from the following relations:
\begin{eqnarray}
U^{(l)}_{s,i}(x) &=& U^{(l)}_{-i,-s}(x), \nonumber \\
U^{(l)}_{s,i}(x) &=& (-1)^{l} U^{(l)}_{s,-i}(x+i) \hspace{5.5mm} i=-1,-2, \nonumber\\
U^{(l)}_{s,i}(x) &=& (-1)^{l} U^{(l)}_{-s,i}(x-s)  \hspace{5mm} s=-1,-2.
\label{eqn:relation}
\end{eqnarray}

Although the obtained expression is rather complicated,
this formula  gives exact values of the time-averaged probability for any
initial state and for any position of lattice site.

Let us write briefly the initial state  by $\Psi_{0}=(\psi(-2,0,0),
\cdots,\psi(2,0,0))$, and consider the time-averaged probability summing  over all possible
states $\bar{P}_{\infty}(x) \equiv \sum_{s \in \Gamma} \bar{P}_{\infty}(s,x)$.
Fig.2 shows $\bar{P}_{\infty}(x)$ for three different initial states: (a)
symmetrical case $\Psi_{0}=(\frac{1}{2},\frac{1}{2},\frac{i}{2},\frac{i}{2})$, (b) asymmetrical case
$\Psi_{0}=(1,0,0,0)$, and (c) extinction case $\Psi_{0}=(-\frac{1}{2},
\frac{1}{2}$,$\frac{1}{2},-\frac{1}{2})$.
One clearly finds that the walker is localized near the origin for the case (a) and
its shape is almost the same with that in Fig.1. In case of (b), there is a double peak and
it shifts  to the left. In contrast with (a) and (b), no peaks exist in the case of (c).
The probability of finding a walker decreases in inverse proportion to the system size and 
it converges to zero in the limit of $N \rightarrow \infty$.

\begin{figure}[h]
\centering
\includegraphics{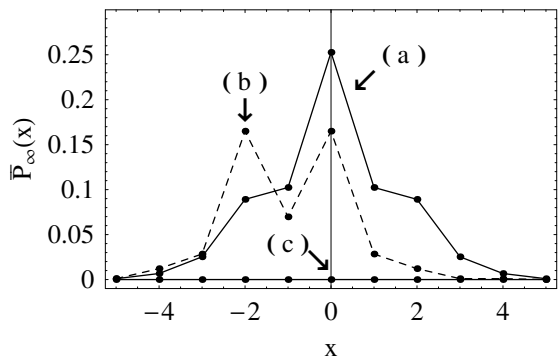}
\caption{\label{fig:Figure3} 
Dependence of the time-averaged probability distribution of the four-state  quantum walk on the initial states $\Psi_{0}$:
(a) ($\frac{1}{2}$,$\frac{1}{2}$,$\frac{i}{2}$,$\frac{i}{2}$), (b) (1,0,0,0), and (c) (-$\frac{1}{2}$,
$\frac{1}{2}$,$\frac{1}{2}$,-$\frac{1}{2}$).
}
\end{figure}

\begin{figure}[h]
\centering
\includegraphics{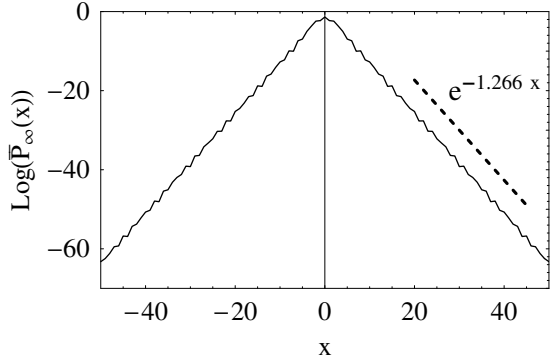}
\caption{\label{fig:Figure4} 
Semi-log plot of the time-averaged probability of four-state quantum walk with initial state (a).
The dashed straight line in the figure shows that  the time-averaged probability decays exponentially  for large $x$.
}
\end{figure}

We move our attention to the asymptotic behavior for large $|x|$. From  Eq. (\ref{eqn:gG})
each component of the matrices $U^{(l)}$ exponentially decreases as increasing the $|x|$.
The values of $\alpha$ and $\beta$ are nearly equal to $-0.381966$ and $-0.53101$, respectively, 
thus the latter contributes largely to $P_{\infty}(x)$.
Fig.3 shows $\log_{e}\bar{P}_{\infty}(x)$ as function of $x$ with the initial state (a). The slope of inserted line
means the exponent of decay is $-1.266$.
This exponential decay of the time-averaged probability $\bar{P}_{\infty}(x)$ shows surely 
that four-state quantum walker localizes near the origin
except for special initial states. 
Very recently similar exponential decay is observed in \it momentum space \rm of generalized quantum walk by Romanelli et al. [12].
We should remark that only the four-state quantum walk starting from 
$\Psi_{0}=\{ -e^{i\varphi},e^{i \varphi},e^{i \varphi},-e^{i \varphi} \}/2$ for any real number $\varphi$ does not show the
localization. This means that the localization is  substantially universal property in four-state quantum walk.

\section{Three-state quantum walk}

We have shown that the four-state quantum walker is fixed near the initial position except special initial conditions.
In this section, three-state quantum walk with asymmetrical jump is considered. The time evolution of 
three-state quantum walk is defined by
\begin{eqnarray}
\psi(-1,x,t+1) &=& \frac{1}{3} \left\{
-\psi(-1,x+1,t) 
+2\psi(1,x+1,t)
+2\psi(2, x+1,t) \right\}
, \nonumber\\ 
\hspace{-15mm}\psi(1,x,t+1) &=&
 \frac{1}{3} \left\{
2\psi(-1,x-1,t) 
-\psi(1,x-1,t)
+2\psi(2, x-1,t)
 \right\}
,
  \nonumber \\ 
\hspace{-15mm}\psi(2,x,t+1) &=&  
\frac{1}{3} \left\{
2\psi(-1,x-2,t)+ 
2\psi(1,x-2,t)
-\psi(2, x-2,t) \right\}. \nonumber\\
\label{eqn:Hdevo}
\end{eqnarray}

In four-state quantum walk, the walker can jump symmetrically to the right  and the left.
This three-state quantum walker, by contrast, can jump to  nearest neighbor sites and  a right-hand  next-nearest neighbor site,
but the walker can not jump to a left-hand  next-nearest neighbor site.

Fig.4 shows snapshots of probability distribution at $t=100,200$ and 300 starting from the origin with an initial state 
$\psi(-2,0,0)=\frac{i}{3}$, $\psi(-1,0,0)=\psi(1,0,0)=\frac{1}{3}$.
We observe that  the several local peaks  move to the right and the heights gradually decrease in 
the prolonged simulation.
\begin{figure}[h]
\centering
\includegraphics{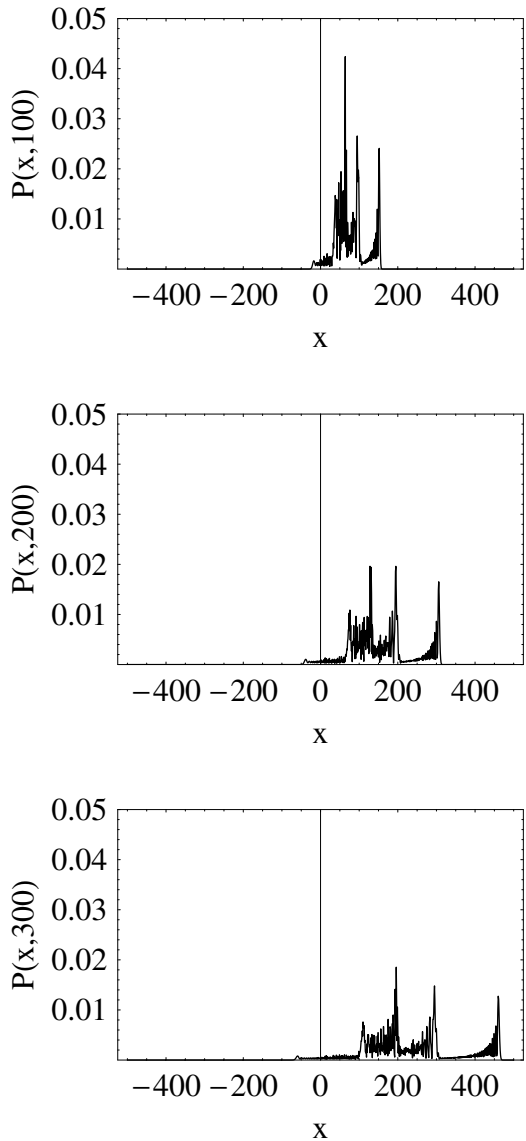}
\caption{\label{fig:Figure4} 
Probability distributions of three-state quantum walk at $t=100,200$ and 300. A spike, which exits initially at the origin,
is divided into several packets and each packet moves to the right.  The height of each peak becomes low as time passes and approaches to zero 
in the limit of $t \rightarrow \infty$.  
}
\end{figure}

The striking difference between four-state  and three-state quantum walks is explained from
difference in  the matrix $H_{k}$. The matrix $H_{k}$ corresponding to three-state quantum walk is given by
\begin{eqnarray}
H_{k}^{(3)}=
\frac{1}{3}
\left[
\begin{array}{ccc}
\omega^{-k} &     0        &     0        \\
0           &  \omega^{k}  &     0        \\
0           &     0        &  \omega^{2k}   \\
\end{array}
\right]
\left[
\begin{array}{ccc}
-1 &  2   &   2 \\
2  & -1   &   2 \\
2  &  2   &  -1 \\ 
\end{array}
\right].
\label{H}
\end{eqnarray}
The eigenvalues of $H_{k}^{(3)}$ are exactly obtained. It is easily found that
the each degree of degeneracy is independent of the system size.
In other words, there is not any eigenvalue corresponding $-1$ and $1$ considered in the previous section.
Therefore the height of each peak inevitably approach as to zero in the thermodynamic limit. Consequently, the localization can  not be
observed in the three-state quantum walk.

\section{Summary}

The key factor of the localization in the four-state quantum walk is the degeneracy of eigenvalues. If the degree of degeneracy is independent of
the dimension of the Hilbert space, the contribution to the time-averaged probability approaches to zero in the limit of 
$N \rightarrow \infty$.  Therefore the existence of eigenvalues whose degree of degeneracy is proportion to the size of the Hilbert space
is a necessary condition. This basic condition is  also satisfied  in the two-dimensional Grover walk.
As a result, the walk shows localization [8].
From this criterion we conclude that the Hadamard walk does not exhibit localization.
Furthermore we  proved that three-state quantum walk with asymmetrical jump 
does not show the localization.

The time-averaged probability at the initial position in  the Grover walk already calculated. However the probability distribution
has not been calculated yet.  Thus we showed  firstly that the time-averaged probability of finding the particle decreases exponentially (not Gaussian)   
as distance increases from a center of a spike and there are special initial states with which the spike disappears in
a one-dimensional quantum walk.

The primary criterion of being localization in probability
distribution is only the dependence of degeneracy of eigenvalue on the system size. If the number of the internal state is even, 
then eigenvalues include $-1$ and $1$ and their degree of degeneracy increase as the system size. Accordingly we can conclude that the multi-state quantum walker
with even internal states almost freezes without detailed calculations.

\end{document}